\documentclass[11pt]{extarticle}
\usepackage[paperheight=12in,paperwidth=8.75in]{geometry}
\usepackage{setspace}
\usepackage[T1]{fontenc}
\usepackage[utf8]{inputenc}
\usepackage[english]{babel}
\usepackage{lmodern}
\usepackage{graphicx}
\usepackage{dcolumn}
\usepackage{amsmath,amssymb,amsfonts}
\usepackage[all]{xy}
\usepackage{cite}
\usepackage{physics}
\usepackage{mathrsfs}
\usepackage{authblk}
\usepackage{abstract}
\usepackage[hidelinks]{hyperref}

\begin{document}
\numberwithin{equation}{section}
\raggedbottom
\onecolumn

\begin{center}
{\Large \bf Quadratic symmetry algebras and Sturm algebraization of superintegrable systems with magnetic fields \\
}
\vspace{6mm}
{Shams Ara$^a$, Md Fazlul Hoque$^a$ and Ian Marquette$^b$ 
}
\\[6mm]
\noindent ${}^{a}${\em 
Pabna University of Science and Technology, Faculty of Science, Department of Mathematics, Pabna 6600, Bangladesh}
\\[3mm]
\noindent ${}^{b}${\em
Department of Mathematical and Physical Sciences, La Trobe University, Bendigo, VIC 3552,
Australia}\\ [1mm]
\vspace{4mm}
{\footnotesize Email: shams.tithi@gmail.com; fazlulmath@pust.ac.bd; i.marquette@latrobe.edu.au }
\end{center}

\vskip 1cm
\begin{abstract}
\noindent We study two three-dimensional quantum superintegrable systems in nonvanishing axially symmetric magnetic fields.  For each model the integrals of motion generate a quadratic algebra with a Casimir operator, and finite-dimensional deformed-oscillator representations yield algebraic energy spectra.  The same spectra are recovered independently by separation of variables.  In circular parabolic coordinates the separated equations are of sextic, biconfluent-Heun type and possess finite polynomial sectors.  We formulate these equations as Sturm eigenvalue problems in which the physical energy is kept as a parameter and the separation constant is diagonalized.  After a gauge transformation, each separated equation admits an $\mathfrak{sl}(2)$ algebraization and reduces to an explicit finite tridiagonal matrix.  We then construct a Sturm representation of each complete Hamiltonian before separation by taking the inverse-square coupling as the Sturm eigenvalue.  The transformed integrals remain commuting operators and, after fixing the axial integral, changing variables and gauge rotating, the full two-variable Sturm Hamiltonian and its non-central integrals belong to the universal enveloping algebra $U(\mathfrak{sl}(3))$.  The analysis therefore provides three complementary algebraic descriptions of the spectrum: the quadratic-algebra/deformed-oscillator representation of the original Hamiltonian, the $\mathfrak{sl}(2)$ algebraization after parabolic separation, and the $\mathfrak{sl}(3)$ algebraization of the complete Sturm problem before separation.  Although the two magnetic Hamiltonians are distinct members of the original classification, fixing the axial integral reduces them to the same singular $2{:}1$ oscillator normal form with different effective parameter embeddings. This explains the common $\mathfrak{sl}(3)$ structure without implying equivalence of the original magnetic systems.  This illustrates how distinct hidden algebraic structures, together with exact and quasi-exact solvability, can emerge at different stages of the analysis of a multiseparable superintegrable system.
\end{abstract}

\section{Introduction}
Superintegrable quantum systems occupy a special position between spectral theory, symmetry analysis and the theory of exactly solvable differential equations.  In addition to the Hamiltonian, they possess more integrals of motion than are required for Liouville integrability, and these additional operators strongly constrain the spectrum, degeneracies and separated wavefunctions.  Familiar examples such as the Coulomb and oscillator systems are governed by Lie symmetries, but more general superintegrable systems are naturally described by polynomial associative algebras generated by their integrals of motion \cite{Mil2013}.  The situation becomes more restrictive in the presence of a magnetic field.  The canonical momenta are replaced by gauge-covariant momenta, the components of the latter no longer commute, and the compatibility conditions for additional integrals involve both the scalar and vector potentials.  Consequently, genuinely superintegrable systems in nonvanishing magnetic fields form a comparatively small class, especially in three dimensions \cite{Marc2015,Marc2017,Br1}.  Their study is therefore useful both for the classification of integrable electromagnetic systems and as a source of nontrivial quantum models for which several algebraic methods can be compared directly.  Related questions concerning quantization and separation in magnetic fields have also been studied in lower-dimensional settings \cite{Berube2004}.

One of the most effective algebraic methods for quantum superintegrable systems is based on the polynomial algebra generated by the integrals of motion.  When two non-central integrals $X$ and $Y$ close quadratically, with $W=[X,Y]$, the resulting associative algebra $Q(3)$ possesses a Casimir operator whose value can be expressed in terms of the central quantities, usually the Hamiltonian and any additional commuting first-order integral.  The representation problem can then be converted into a deformed-oscillator problem following the construction of Daskaloyannis \cite{Dask1}.  In its standard form one introduces a number operator $N$ and creation and annihilation operators $b^{\dagger}$ and $b$ satisfying
\[
[N,b^{\dagger}]=b^{\dagger},\qquad [N,b]=-b,\qquad
b^{\dagger}b=\Phi(N),\qquad bb^{\dagger}=\Phi(N+1),
\]
where the structure function $\Phi$ depends on the central elements of the polynomial algebra and therefore on the energy.  A finite-dimensional unitary representation of dimension $p+1$ is obtained from
\[
\Phi(0)=0,\qquad \Phi(p+1)=0,\qquad
\Phi(n)>0\quad (n=1,\ldots,p).
\]
This is achieved by a realization of $X$, $Y$ and $W$ in terms of $N$, $b$ and $b^{\dagger}$. The first two conditions impose algebraic restrictions on the energy and on the shift parameter entering the oscillator realization, while the positivity condition selects the admissible branches.  In this way the spectral problem is reduced to finite-dimensional representation theory.  This approach is particularly useful when a direct solution of the Schr\"odinger equation is possible as well, because the algebraic spectrum can then be checked against separation of variables and the corresponding wavefunctions.  Variants of this quadratic-algebra/deformed-oscillator strategy have been used for $N$-dimensional Kepler--Coulomb and singular-oscillator systems and for superintegrable systems with Abelian and Yang--Coulomb monopole interactions \cite{Hoque2015N, Hoque2015Double, Hoque2016Taub, Hoque2017Yang}.  These applications of the deformed-oscillator method are especially relevant here because the present problem also involves a three-generator quadratic algebra in three dimensions. Monopole systems provide a broader setting for hidden symmetries, Runge--Lenz operators and algebraic spectral constructions.  Alongside the MICZ--Kepler systems of Zwanziger and McIntosh-Cisneros \cite{Zwan1, Mcin1}, relevant developments include Yang's $SU(2)$ monopole \cite{Yang1978}, Jackiw's monopole dynamical symmetry \cite{Jackiw1980}, and the dynamical supersymmetry of monopole systems with an inverse-square interaction studied by D'Hoker and Vinet \cite{DHokerVinet1985}.

Further examples include the monopole-scattering symmetry of Feh\'er and Horv\'athy \cite{FeherHorvathy1987}, the hidden $SO(6)$ symmetry of the Yang--Coulomb monopole \cite{Mardoyan1999}, and the generalized MIC--Kepler system, which is separable in spherical and parabolic coordinates \cite{Mardoyan2003}. The reader can also get further references on the all-dimensional MICZ treatment of Meng \cite{Meng1}.  These works provide useful background for the present combination of magnetic fields, additional integrals and algebraic spectral methods. The systems considered here are not monopole models, and they involve nonuniform axially symmetric magnetic fields for which algebraic spectral derivations remain comparatively little explored.

A complementary viewpoint is provided by quasi-exact solvability.  A differential operator is quasi-exactly solvable when it preserves a finite-dimensional space of functions, so that a finite part of its spectrum can be obtained algebraically.  In one variable, after a suitable change of variable and gauge transformation, the standard Lie-algebraic realization is generated by $\mathfrak{sl}(2)$ acting on the polynomial module $\mathcal P_N=\langle 1,x,\ldots,x^N\rangle$, and the sextic oscillator is one of the canonical examples of this mechanism \cite{turbiner}.  In two variables, polynomial modules such as $\mathcal P_N^{(2)}= \langle x^i y^j\,;\ i+j\leq N\rangle$ are naturally associated with projective realizations of $\mathfrak{sl}(3)$ \cite{Turbiner1992}. Most Lie-algebraic searches for hidden symmetries have, however, been carried out for scalar interactions without magnetic fields. These Lie-algebraic descriptions need not replace the polynomial symmetry algebra of a superintegrable Hamiltonian: they characterize a different aspect of the same spectral problem, namely the invariant polynomial spaces of the gauge-rotated differential operators.

The Sturm formulation gives a useful bridge between these viewpoints.  Instead of always treating the physical energy as the eigenvalue, one may keep $E$ fixed and promote another parameter, for example a coupling constant or a separation constant, to the spectral variable.  Such a reorganization can expose an algebraic structure that is less transparent in the original Schr\"odinger representation.  A recent example is the Sturm-algebraic treatment of a matrix Coulomb-type Hamiltonian in which the integrals are first obtained in Sturm representation and then related back to those of the Schr\"odinger problem \cite{Quesne2026}.  In the present work we use the same general philosophy, but for scalar Hamiltonians in nonuniform magnetic fields: the separated parabolic equations lead to one-variable $\mathfrak{sl}(2)$ algebraizations, while the complete two-variable Sturm problems admit $\mathfrak{sl}(3)$ realizations. One purpose of this paper is to connect these three viewpoints and clarify the distinct algebraic structures that they reveal. 

Against this background of complementary algebraic approaches, we consider two three-dimensional axially symmetric superintegrable systems in magnetic fields obtained in Ref.~\cite{Br1}.  The two Hamiltonians involve different nonuniform magnetic fields and different realizations of the conserved integrals.  Nevertheless, we will show that their quadratic algebras belong to the same general class. We will treat the two models separately at the level of explicit formulas while using the same sequence of calculations: polynomial symmetry algebra, deformed-oscillator spectrum, separation of variables, one-variable $\mathfrak{sl}(2)$ algebraization and, finally, a complete Sturm representation with a two-variable $\mathfrak{sl}(3)$ realization.  This parallel treatment allows the algebraic results to be checked against direct differential-equation calculations rather than relying on a single method.

The organization of the paper is as follows.  In Secs.~2 and 3 we determine the integrals, quadratic algebra, Casimir operator, deformed-oscillator structure function and finite-dimensional spectrum of the first model.  Section~4 studies separation in circular parabolic and cylindrical coordinates and incorporates the corresponding $\mathfrak{sl}(2)$ Sturm algebraization in the same section.  Sections~5 and 6 give the parallel quadratic-algebra, spectral, separation and $\mathfrak{sl}(2)$ analysis for the second magnetic system.  In Sec.~7 we return to the complete Hamiltonians before separation, promote the inverse-square coupling to a Sturm eigenvalue, transform the commuting integrals accordingly and show that, after fixing the axial integral and gauge rotating, the resulting two-variable operators can be expressed in $U(\mathfrak{sl}(3))$.  The final section summarizes how the quadratic algebra, deformed oscillator, exact separation and $\mathfrak{sl}(2)$/$\mathfrak{sl}(3)$ algebraizations fit together.  This organization makes it possible to distinguish model-specific magnetic data from algebraic structures shared by the two systems and to compare exact and quasi-exact solvability before and after separation. 

\section{Integrals of motion and quadratic algebra}

We begin with the Hamiltonian obtained in Ref.~\cite{Br1}
\begin{eqnarray}
H=\frac{1}{2}\left(\mathbf p^A\right)^2-\frac{r^2}{32}(2b_1+b_2(r^2+4z^2))^2+u_1 z+\frac{u_2}{r^2}+u_3(r^2+4z^2),\label{Hamil1}
\end{eqnarray}
moving in the non-uniform axially symmetric magnetic field
\begin{eqnarray}
\mathbf B=(-2b_2xz,-2b_2yz,b_1+b_2(r^2+2z^2)),\label{mg1}
\end{eqnarray}
with additional scalar interactions.  Here $r$ is the cylindrical radius, $r^2=x^2+y^2$, the gauge-covariant momenta are $p_i^A=p_i+A_i$, with $p_i=-i\hbar\partial/\partial x_i$, and we use the vector potential
\begin{eqnarray}
A_x=-\left( \frac{b_1}{2}+\frac{b_2}{4}(r^2+4z^2)\right)y,\quad A_y=\left( \frac{b_1}{2}+\frac{b_2}{4}(r^2+4z^2)\right)x,\quad A_z=0.
\end{eqnarray}
The parameter $b_1$ produces a uniform contribution to the magnetic field, whereas $b_2$ produces a spatially varying contribution.  Magnetic fields with nonzero gradients occur naturally in magnetic confinement, mirror and trap configurations and charged-particle optics \cite{Jack1,Nort1,Brow1,Frei1,Hump1}. In our case, the field is considered primarily as part of a finely tuned superintegrable quantum system.

An important feature of the Hamiltonian is the compensating scalar interaction proportional to $r^2(2b_1+b_2(r^2+4z^2))^2$.  For the vector potential above, this term is precisely the negative of the diamagnetic contribution $A^2/2$ generated by minimal coupling.  The quadratic vector-potential contribution is therefore cancelled, while the linear magnetic coupling survives.  This fine tuning is one of the mechanisms that keeps the differential equations and symmetry operators sufficiently structured to permit exact and quasi-exact treatments.  Related symmetry-restoring and fine-tuning mechanisms occur in other superintegrable constructions, including MICZ-type systems and singular oscillator families \cite{Zwan1,Mcin1,Mardoyan2003,Meng1,Maka1,Evan1}.

The uniform-field limit illustrates why it is useful to distinguish the magnetic coupling from the ordinary Landau Hamiltonian \cite{Land1}.  If $b_2=u_1=u_2=u_3=0$, then
\[
\mathbf A=\left(-\frac{b_1y}{2},\frac{b_1x}{2},0\right),\qquad
H_{\rm u}=\frac12(\mathbf p+\mathbf A)^2-\frac{b_1^2r^2}{8}
=\frac12p^2+\frac{b_1}{2}L_z.
\]
The magnetic field is the uniform Landau field $\mathbf B=(0,0,b_1)$, but the compensating scalar interaction removes the usual diamagnetic term.  When $b_2=0$ while $u_3$ is retained, the transverse motion is of Fock--Darwin type, supplemented in general by the linear and inverse-square interactions \cite{Fock1,Darw1}.  The term $u_2/r^2$ is the familiar singular interaction occurring in conformal quantum mechanics and Calogero-type systems \cite{Calo1,Alfa1}. In the present models it also enters directly into the representation parameters of the quadratic and Lie-algebraic constructions.

We first analyze the symmetry algebra of the first magnetic model before turning to the differential equation.  This order is useful because closure of the commutators determines whether the conserved quantities form a finite polynomial algebra.  When such a closure exists for the integrals and their commutators, the spectrum can be extracted from finite-dimensional representations.
Besides the Hamiltonian (\ref{Hamil1}), the system admits the following three integrals of motion:
\begin{eqnarray}
&&X=(p_z^A)^2+2b_2z^2L_z^A-\left(b_1b_2+\frac{b_2^2}{2}(r^2+4z^2)\right)z^2r^2+2u_1z+8u_3z^2,
\\&&
Y=\frac{1}{2}( L_x^Ap_y^A+p_y^AL_x^A)-\frac{1}{2}(L_y^A p_x^A+p_x^A L_y^A)+(b_1+b_2(r^2+2 z^2))zL_z^A-\frac{b_1^2}{4}zr^2\nonumber\\&& -\frac{b_1b_2}{2}zr^2(r^2+2z^2)-\frac{b_2^2}{16}zr^2(3r^2+4z^2)(r^2+4z^2)+\frac{u_1}{2}r^2-\frac{2u_2 z}{r^2}+2u_3 zr^2,
\\&&
Z=L_z^A-\left( \frac{b_1}{2}+\frac{b_2}{4}(r^2+4z^2)\right)r^2.
\end{eqnarray}
The operators $X$ and $Y$ are second-order integrals, while $Z$ is the axial first-order integral.  They satisfy the commutation relations
\begin{eqnarray}
[H,X]=0=[H,Y], \quad [H,Z]=[X,Z]=[Y,Z]=0.
\end{eqnarray}
Thus $Z$ commutes not only with $H$ but also with the two non-central integrals $X$ and $Y$. It therefore acts as a central element in the symmetry algebra.  Its role should be distinguished from an external central parameter such as a monopole charge: here $Z$ is itself a genuine first-order integral of the three-dimensional Hamiltonian.  The pattern of commuting and non-commuting generators can be summarized by the following diagram.

\begin{eqnarray}
\begin{xy}
(0,0)*+{X}="m"; (20,0)*+{Z}="q"; (40,0)*+{Y}="p"; (20,20)*+{H}="r"; 
"r";"p"**\dir{--};
"m";"q"**\dir{--}; % corrected: commuting edge X--Z (was X--Y)
"q";"p"**\dir{--}; 
"r";"m"**\dir{--};
"r";"q"**\dir{--};
\end{xy}
\end{eqnarray}

The remaining nontrivial commutators close polynomially.  This closure is the key algebraic property used in the next section: it replaces an infinite commutator hierarchy by a finitely generated associative algebra.  Defining $W=[X,Y]$, one obtains the quadratic algebra $Q(3)$ given by the commutator relations below
\begin{eqnarray}
&&[X,Y]=W,\nonumber
\\&&
[X,W]=4 \hbar^2 u_1 X+8\hbar^2(b_2 Z+4u_3)Y-8\hbar^2 u_1 H+4 \hbar^2 b_1 u_1 Z,\label{QA1}
\\&&
[Y,W]=6\hbar^2 X^2+8\hbar^2(b_1 Z-2 H)X-4\hbar^2 u_1 Y+8\hbar^2 H^2 - 8\hbar^2 b_1 Z H - 4\hbar^2 b_2 Z^3\nonumber\\&&\qquad\qquad +2\hbar^2(b_1^2-8 u_3)Z^2 +4\hbar^2(\hbar^2-2u_2)(b_2 Z+4 u_3).\nonumber
\end{eqnarray}
The central elements $H$ and $Z$ enter only through the structure constants.  Although these structure constants contain powers as high as $Z^3$, the algebra is quadratic in the three non-Abelian generators $X$, $Y$ and $W$.  It can be written in the standard form
\begin{equation}
[X,Y]=W,\quad [X,W]=\delta X + \epsilon Y + \zeta,\quad [Y,W]=a X^2 +d X - \delta Y + z
\end{equation}
where the structure constants are given by
\begin{align}
\delta&=4\hbar^2u_1, \qquad \epsilon= 8\hbar^2(b_2Z+4u_3),\nonumber\\
\zeta&=4\hbar^2u_1(b_1Z-2H),\qquad
a=6\hbar^2,\qquad d=8\hbar^2(b_1Z-2H),\nonumber\\
z&=8\hbar^2H^2-8\hbar^2b_1ZH-4\hbar^2b_2Z^3
+2\hbar^2(b_1^2-8u_3)Z^2\nonumber\\
&\qquad+4\hbar^2b_2(\hbar^2-2u_2)Z
+16\hbar^2u_3(\hbar^2-2u_2).
\end{align}

The construction of the deformed-oscillator algebra requires the Casimir operator.  Its role is twofold: it provides an internal check of the closure relations and supplies the central quantity needed in the deformed-oscillator realization. In the present normalization it is
\begin{eqnarray}
&& K=W^2-4\hbar^2 u_1\{X,Y\}-8\hbar^2(b_2 Z+4u_3)Y^2-8\hbar^2 u_1(b_1 Z-2H)Y+4\hbar^2 X^3 \nonumber\\&&\qquad +8\hbar^2(b_1 Z-2H)X^2+\left(16\hbar^2 H^2-16\hbar^2 b_1 Z H-8\hbar^2 b_2 Z^3+4\hbar^2 (b_1^2-8u_3)Z^2\right. \nonumber\\&&\qquad \left. +8\hbar^2 b_2(3\hbar^2 -2u_2)Z+32\hbar^2 u_3(3\hbar^2-2 u_2)\right)X.
\end{eqnarray}

By construction, this Casimir commutes with $X$, $Y$ and $W$.  Substituting the differential realization of the integrals simplifies it considerably: the result depends only on the central elements $H$ and $Z$, namely
\begin{eqnarray}
&& K=-4\hbar^2 (4\hbar^2 b_1 b_2 - u_1^2)Z^2+32\hbar^4 b_2 H Z+128\hbar^4 u_3 H - 64\hbar^4 b_1 u_3 Z\nonumber\\&& \qquad\quad -4\hbar^2 u_1^2(\hbar^2 - 2 u_2).\label{Casimir2}
\end{eqnarray} 

\section{Structure function and energy spectrum}
We now use the quadratic algebra to derive the spectrum algebraically.  It is useful first to make explicit how the quadratic algebra is embedded in the deformed-oscillator algebra introduced in Sec.~1, namely $[\mathcal N,b^\dagger]=b^\dagger$, $[\mathcal N,b]=-b$, $b^\dagger b=\phi(\mathcal N)$ and $b b^\dagger=\phi(\mathcal N+1)$. The realization \cite{Dask1} can therefore be written in the form
\[
X=\sqrt{\epsilon}\,(\mathcal N+q),\qquad
Y=\beta(\mathcal N)+b^\dagger\rho(\mathcal N)+\rho(\mathcal N)b,
\qquad
W=\sqrt{\epsilon}\,[b^\dagger\rho(\mathcal N)-\rho(\mathcal N)b],
\]
\[
\beta(\mathcal N)=-\frac{\delta}{\sqrt{\epsilon}}(\mathcal N+q)-\frac{\zeta}{\epsilon}.
\]
Here $q$ is the shift parameter of the oscillator realization.  The first two quadratic-algebra relations are then satisfied identically.  The remaining commutator $[Y,W]$, together with the value of the Casimir, determines the structure function $\phi$.  Thus the structure function used below is not an independent ansatz: it is the quantity that completes a deformed-oscillator realization of the quadratic algebra.  There is also a normalization freedom in $\rho$: the quadratic relations determine the combinations $\phi(\mathcal N+1)\rho^2(\mathcal N)$ and $\phi(\mathcal N)\rho^2(\mathcal N-1)$.  Different choices of $\rho$ therefore give equivalent oscillator realizations but can change the displayed factorization of $\phi$. Using the central Casimir (\ref{Casimir2}), we restrict to a simultaneous eigenspace of the commuting operators $H$ and $Z$, writing $H=E$ and $Z=m\hbar$.  It is convenient to introduce the two scales
\begin{eqnarray}
\Omega^2=2(4u_3+b_2m\hbar),\qquad \Lambda^2=m^2\hbar^2+2u_2.
\label{scalesM1clean}
\end{eqnarray}
Since $\epsilon=4\hbar^2\Omega^2$, the realization specializes to
\begin{align}
X={}&2\hbar\Omega(\mathcal N+q),\nonumber\\
Y={}&-\frac{2\hbar u_1}{\Omega}(\mathcal N+q)
+\frac{u_1(2E-b_1m\hbar)}{\Omega^2}
+b^\dagger\rho_1(\mathcal N)+\rho_1(\mathcal N)b,\nonumber\\
W={}&2\hbar\Omega[b^\dagger\rho_1(\mathcal N)-\rho_1(\mathcal N)b].
\label{M1oscReal}
\end{align}
For the factorized normalization used in this paper we choose
\[
\rho_1(\mathcal N)=\left(\mathcal N+q+1-\eta_1\right)^{-1/2},
\qquad
\eta_1=\frac{\hbar\Omega^3-u_1^2}{2\hbar\Omega^3}.
\]
With the canonical choice $\rho_1=1$ the same quadratic algebra gives an equivalent cubic structure function.  The above normalization simply absorbs one linear factor into the oscillator ladders and yields the convenient quartic factorization displayed next.  Substitution of (\ref{M1oscReal}) into $[Y,W]$ and the Casimir relation gives
\begin{eqnarray}
\phi(t,E,q) = && 2\hbar^3 \Omega\left (t+q-\frac{ \hbar \Omega^3 - u_1^2 }{2\hbar \Omega^3 } \right)^2\left (t+q-\frac{2 E +  \hbar \Omega - \Omega\Lambda -\hbar  b_1  m}{2\hbar \Omega } \right)\nonumber\\&&\times\left (t+q-\frac{2 E +  \hbar \Omega + \Omega\Lambda -\hbar  b_1  m}{2\hbar \Omega } \right).
\label{M1factorPhi}
\end{eqnarray}
This factorized expression is the useful form for representation theory because the endpoints of a finite deformed-oscillator module can be read off from its zeros.
For a finite-dimensional unitary representation of dimension $p+1$, with $p=0,1,2,\ldots$, the structure function must vanish at the two endpoints and be positive at the intermediate integer values.  When the factorized normalization specified above is real and nonsingular on the module, we therefore impose
\begin{eqnarray}
\phi(0,E,q)=0,\quad \phi(p+1,E,q)=0,\quad \phi(t)>0,\quad t=1,\ldots,p.
\end{eqnarray}
Solving the two endpoint conditions gives the following algebraic branches for the shift $q$ and the energy, with $\varepsilon_1,\varepsilon_2=\pm1$.  For the first branch,
\begin{align}
E_1&=\frac{1}{2}\left(b_1 \hbar m + 2\hbar \Omega(p+1)+ \varepsilon_1 \Omega\Lambda-\frac{u_1^2}{\Omega^2}\right),\label{E1plusBranch}\\
\phi_1(t)&=2\hbar^2\Omega t^2(p+1-t)\bigl[\hbar(p+1-t)+\varepsilon_1\Lambda\bigr], \quad q_1=\frac{1}{2}\left(1-\frac{u_1^2}{\hbar \Omega^3}\right).
\end{align}
For the second branch,
\begin{align}
E_2&=\frac{1}{2}\left(b_1\hbar m-2\hbar\Omega(p+1)+\varepsilon_2\Omega\Lambda-\frac{u_1^2}{\Omega^2}\right),\\
\phi_2(t)&=2\hbar^2\Omega t(\hbar t-\varepsilon_2\Lambda)(p+1-t)^2, \quad q_2=\frac{1}{2}-\frac{b_1\hbar m-2E_2+\varepsilon_2\Omega\Lambda}{2\hbar\Omega}.
\end{align}

For the first branch, $q_1=\eta_1$, and hence $\rho_1^2(\mathcal N)=1/(\mathcal N+1)>0$.  Thus the standard positivity condition applies directly to $\phi_1(t)$.  In particular, for $\Omega>0$ the regular positive-Frobenius choice $\varepsilon_1=+1$ gives the physical branch, which will be recovered independently below from both cylindrical separation and the Sturm $\mathfrak{sl}(2)$ construction.  The second branch is retained as an additional algebraic solution of the endpoint conditions but is not used below to identify the regular physical spectrum.

\section{Separation of variables for the first model}
The spectral information obtained from the quadratic algebra should be checked against a direct solution of the Schr\"{o}dinger equation.  We therefore separate the same Hamiltonian in two complementary coordinate systems.  Circular parabolic coordinates expose the sextic equations that underlie the quasi-exactly solvable $\mathfrak{sl}(2)$ sector and determine the parabolic separation constants.  Cylindrical coordinates, by contrast, reduce the problem to a singular radial oscillator and a shifted one-dimensional oscillator, so that the normalizable wavefunctions and energy quantization can be written explicitly in standard special functions.  Agreement between the two separations and the algebraic spectrum provides an important consistency check.
\subsection{Parabolic coordinates}
We consider the circular parabolic coordinates
\[
x=\xi \eta \cos\phi,\qquad y=\xi \eta \sin\phi,\qquad z=\frac{1}{2}(\xi^2-\eta^2),
\]
where $\xi,\eta>0$ and $0\leq\phi\leq2\pi$.  In these coordinates the vector potential is
\begin{align*}
A_1&=-\left(\frac{b_1}{2}+\frac{b_2}{4}(\xi^4-\xi^2\eta^2+\eta^4)\right)\xi\eta\sin\phi,\\
A_2&=\left(\frac{b_1}{2}+\frac{b_2}{4}(\xi^4-\xi^2\eta^2+\eta^4)\right)\xi\eta\cos\phi,\\
A_3&=0.
\end{align*}
The Schr\"{o}dinger equation $H\Psi=E\Psi$ then takes the form
\begin{eqnarray}
&& -\frac{\hbar}{\xi^2+\eta^2}\left[2\hbar\xi^2\eta^2\left(\frac{\partial^2 \Psi}{\partial \xi^2}+\frac{\partial^2 \Psi}{\partial \eta^2}\right)+2\hbar\xi\eta\left(\eta\frac{\partial \Psi}{\partial \xi}+\xi\frac{\partial \Psi}{\partial \eta}\right)+2\hbar (\xi^2+\eta^2)\frac{\partial^2 \Psi}{\partial \phi^2}\right.\nonumber\\&& \left.\qquad +(2 i b_1\xi^2\eta^2(\xi^2+\eta^2)+i b_2\xi^2\eta^2(\xi^6+\eta^6))\frac{\partial \Psi}{\partial \phi}\right]+[4u_2-4E\xi^2\eta^2 +2u_1\xi^2\eta^2(\xi^2-\eta^2)\nonumber\\&& \qquad +4u_3 \xi^2\eta^2(\xi^4-\xi^2\eta^2+\eta^4)]\Psi=0.
\label{Schr1}
\end{eqnarray}
Substituting the separated ansatz
\[
\Psi(\xi,\eta,\phi)=f_1(\xi)f_2(\eta)e^{im\phi}
\]
into Eq.~(\ref{Schr1}) gives two ordinary differential equations.  To display their common structure, set $\varsigma=b_2m/(2\hbar)+2u_3/\hbar^2$, $v_1=u_1/\hbar^2$, $v_2=-v_1$, $\nu=b_1m/\hbar-2E/\hbar^2$, and $\sigma=m^2+2u_2/\hbar^2$.  If $\kappa_1$ denotes the scaled separation constant, then $\kappa_2=-\kappa_1$, and both equations take the form
\begin{eqnarray}
\frac{d^2f_i}{dz_i^2}+\frac1{z_i}\frac{df_i}{dz_i}
-\left(\varsigma z_i^6+v_i z_i^4+\nu z_i^2+\frac{\sigma}{z_i^2}-\kappa_i\right)f_i=0,
\qquad z_1=\xi,\quad z_2=\eta.
\label{conf3}
\end{eqnarray}
The opposite signs $v_2=-v_1$ and $\kappa_2=-\kappa_1$ are the two features that distinguish the $\xi$ and $\eta$ equations.  Otherwise they are the same sextic singular oscillator.  The inverse-square term fixes the local Frobenius behavior at the origin, whereas the sextic term controls the large-$z_i$ decay.  Factoring out these two asymptotic contributions is what reduces the remaining equation to biconfluent-Heun form. To connect this equation with the biconfluent Heun equation, we set $\chi_i=z_i^2$. Let us remark that useful properties and alternative representations of this equation are discussed, for example, in Ref.~\cite{Ish1}.  Equation (\ref{conf3}) becomes
\[
4\chi_i\frac{d^2f_i}{d\chi_i^2}+4\frac{df_i}{d\chi_i}
-\left(\varsigma\chi_i^3+v_i\chi_i^2+\nu\chi_i+\frac{\sigma}{\chi_i}-\kappa_i\right)f_i=0.
\]
We extract the behavior at the origin and at infinity by writing
\[
f_i(\chi_i)=\chi_i^\alpha\exp[-(\beta\chi_i^2+\mu_i\chi_i)]P_i(\chi_i),
\]
where $\alpha=\sqrt\sigma/2$, $\beta=\sqrt\varsigma/4$, and $\mu_i=v_i/(4\sqrt\varsigma)$.
The reduced equation is
\begin{eqnarray}
P_i''+\left(\frac{2\alpha+1}{\chi_i}-4\beta\chi_i-2\mu_i\right)P_i'
+\left(A_i+\frac{B_i}{\chi_i}\right)P_i=0,
\label{BHEredclean}
\end{eqnarray}
where $A_i=\mu_i^2-4\beta(\alpha+1)-\nu/4$ and $B_i=\kappa_i/4-\mu_i(2\alpha+1)$.  Comparing Eq.~(\ref{BHEredclean}) with the canonical biconfluent-Heun equation
\[
y''+\left(\frac{\lambda_1+1}{x}-\lambda_2-\lambda_3x\right)y'
+\left(\frac{\lambda_4x-\lambda_5}{x}\right)y=0
\]
gives $\lambda_1=2\alpha$, $\lambda_2=2\mu_i$, $\lambda_3=4\beta$, $\lambda_4=A_i$, and $\lambda_5=-B_i$.  Thus a local solution may be written as
\[
f_i(z_i)=z_i^{2\alpha}
\exp\left[-\frac{\sqrt\varsigma}{4}z_i^4-\frac{v_i}{4\sqrt\varsigma}z_i^2\right]
\mathrm{HeunB}(\lambda_1,\lambda_2,\lambda_3,\lambda_4,\lambda_5;z_i^2).
\]
Here $\mathrm{HeunB}$ denotes the local biconfluent-Heun solution in the normalization of Eq.~(\ref{BHEredclean}).  A generic Heun solution is an infinite series.  The quasi-exact polynomial sector appears only when two independent termination conditions are satisfied.
For a polynomial of degree $N$, cancellation of the highest power in Eq.~(\ref{BHEredclean}) gives
\begin{equation}
A_i=N\sqrt\varsigma,
\qquad\text{equivalently}\qquad
\nu=\frac{v_i^2}{4\varsigma}-4\sqrt\varsigma\,(N+\alpha+1).
\label{HeunEnergyCondClean}
\end{equation}
A second termination condition is required to stop the coefficient recurrence. It is then a polynomial equation of degree $N+1$ for the separation constant $\kappa_i$.  The final subsection of the present section derives this condition explicitly from the $\mathfrak{sl}(2)$ representation and obtains the corresponding finite tridiagonal matrix.  This also makes the connection with the quadratic-algebra spectrum transparent.

\subsection{Cylindrical coordinates}
We next turn to cylindrical coordinates.  In this coordinate system the cancellation of the diamagnetic term leaves an effective potential that is quadratic in $\rho$ and $z$ after the azimuthal dependence has been separated.  This makes the exact bound-state problem considerably more transparent.  The Schr\"{o}dinger equation $H\Psi(\rho,\phi,z)=E\Psi(\rho,\phi,z)$ reads
\begin{align}
\Bigg[&-\frac{\hbar^2}{2}\left(\frac{\partial^2}{\partial\rho^2}
+\frac1\rho\frac{\partial}{\partial\rho}
+\frac1{\rho^2}\frac{\partial^2}{\partial\phi^2}
+\frac{\partial^2}{\partial z^2}\right)
-i\hbar\left(\frac{b_1}{2}+\frac{b_2}{4}(\rho^2+4z^2)\right)\frac{\partial}{\partial\phi}\nonumber\\
&\qquad +u_1z+\frac{u_2}{\rho^2}+u_3(\rho^2+4z^2)\Bigg]\Psi=E\Psi.
\label{Cylin1}
\end{align}
The cancellation of the $A^2/2$ term used in Sect.~2 is what makes this cylindrical form quadratic in $\rho$ and $z$ after the azimuthal separation.
 
Using the separated ansatz $\Psi(\rho,\phi,z)=f_1(\rho)f_2(z)e^{im\phi}$, the Schr\"{o}dinger equation (\ref{Cylin1}) separates into
 \begin{eqnarray}
 && \frac{d^2f_1(\rho)}{d\rho^2}+\frac{1}{\rho}\frac{df_1(\rho)}{d\rho}+\left(\tau_0 -\frac{\tau_1}{\rho^2}-\tau_2 \rho^2 \right)f_1(\rho)=0,\label{Req1}\\&&
 \frac{d^2f_2(z)}{d z^2}-\left(\tau_3 z^2+2u'_1 z-k_1\right)f_2(z)=0,\label{zeq1}
 \end{eqnarray}
where $u'_1=\frac{u_1}{\hbar^2}$, $u'_2=\frac{u_2}{\hbar^2}$, $u'_3=\frac{u_3}{\hbar^2}$, $b'_1=\frac{b_1}{\hbar}$, $b'_2=\frac{b_2}{2\hbar}$, $E'=\frac{E}{\hbar^2}$, $\tau_0=2E'-b'_1 m-k_1$, $\tau_1=m^2+2u'_2$, $\tau_2=b'_2 m+2u'_3$, $\tau_3=4(b'_2 m+2u'_3)$ and $k_1$ is the separation constant.    
 The radial equation is a singular oscillator.  Set $\chi=\varepsilon\rho^2$, with $\varepsilon=\sqrt{\tau_2}$, and write $f_1(\chi)=\chi^{\sqrt{\tau_1}/2}e^{-\chi/2}f(\chi)$.  Equation (\ref{Req1}) becomes Kummer's confluent hypergeometric equation
 \begin{eqnarray}
 \chi \frac{d^2f(\chi)}{d\chi^2}+\left((\sqrt{\tau_1}+1)-\chi\right)\frac{df(\chi)}{d\chi} -\left(\frac{\sqrt{\tau_1}+1}{2}-\frac{\tau_0}{4\varepsilon}\right)f(\chi)=0,\label{hypg1}
 \end{eqnarray}
In the standard notation ${}_1F_1(a;b;x)$ denotes Kummer's confluent hypergeometric function of the first kind. The normalizability at large $\rho$ requires the confluent-hypergeometric series to terminate; this implies that its first parameter must be a non-positive integer.  Hence
 \begin{eqnarray}
 n=\frac{\tau_0}{4\varepsilon}-\frac{\sqrt{\tau_1}+1}{2},\label{nn1}
 \end{eqnarray}
and Eq.~(\ref{hypg1}) gives the radial solution
\begin{eqnarray}
 f_1(\rho)\propto \rho^{\sqrt{\tau_1}} e^{-\frac{\sqrt{\tau_2}}{2}\rho^2} {}_1F_1\left(-n;\sqrt{\tau_1}+1;\sqrt{\tau_2}\rho^2 \right).
 \end{eqnarray}
 
Because the first argument is $-n$, the hypergeometric factor is proportional to the generalized Laguerre polynomial $L_n^{(\sqrt{\tau_1})}(\sqrt{\tau_2}\rho^2)$.  Thus $n=0,1,2,\ldots$ is the radial quantum number, and the regular radial solution is square integrable for the positive Frobenius exponent.
The $z$ equation in Eq.~(\ref{zeq1}) is a shifted harmonic oscillator.  Completing the square with $\xi=z+u'_1/\tau_3$ and $k'_1=k_1+{u'_1}^2/\tau_3$, then introducing $t=\tau_3^{1/4}\xi$ and $f_2=e^{-t^2/2}f_3$, gives the Hermite equation
\begin{equation}
\frac{d^2f_3}{dt^2}-2t\frac{df_3}{dt}+\left(\frac{k'_1}{\sqrt{\tau_3}}-1\right)f_3=0.
\label{M1HermiteEq}
\end{equation}
A bounded solution requires $k'_1/\sqrt{\tau_3}-1=2\lambda$ with $\lambda=0,1,2,\ldots$.  Hence
\[
k_1=(2\lambda+1)\sqrt{\tau_3}-\frac{{u'_1}^2}{\tau_3},
\qquad
f_2(z)\propto e^{-\frac{\sqrt{\tau_3}}2\left(z+\frac{u'_1}{\tau_3}\right)^2}
H_\lambda\!\left(\tau_3^{1/4}\left(z+\frac{u'_1}{\tau_3}\right)\right).
\]
Here $H_\lambda$ is the Hermite polynomial of degree $\lambda$, and $\lambda$ is the longitudinal oscillator quantum number.  Combining this condition with the radial quantization condition (\ref{nn1}) gives
\begin{equation}
E'=(2n+\sqrt{\tau_1}+1)\sqrt{\tau_2}+\frac12\left[b'_1m+(2\lambda+1)\sqrt{\tau_3}-\frac{{u'_1}^2}{\tau_3}\right],
\qquad n,\lambda=0,1,2,\ldots.
\label{M1cylEnergy}
\end{equation} 
 
The two non-negative integers $n$ and $\lambda$ label the radial and longitudinal excitations.  Their sum will be identified below with the polynomial degree $N$ of the parabolic $\mathfrak{sl}(2)$ module, which makes the agreement between the exact cylindrical spectrum and the quasi-exact parabolic sector explicit.
\subsection{Sturm form and explicit \texorpdfstring{$\mathfrak{sl}(2)$}{sl(2)} algebraization}
The Heun reduction above already identifies when a finite polynomial sector can occur, but it does not by itself display the finite matrix whose eigenvalues are the allowed separation constants.  It is therefore useful to reorganize the same separated equation as a Sturm eigenvalue problem for $\kappa_i$.  In this formulation the physical energy remains a parameter, the second Heun termination condition becomes an ordinary characteristic equation, and the finite-dimensional $\mathfrak{sl}(2)$ action is explicit.

Recall the scales $\Omega$ and $\Lambda$ introduced in Eq.~(\ref{scalesM1clean}), and set $a=\Omega/(2\hbar)$, $c_1=u_1/(\hbar\Omega)$, $c_2=-c_1$, and $s=\varepsilon\Lambda/(2\hbar)$ with $\varepsilon=\pm1$.  For $z_1=\xi$ and $z_2=\eta$, Eq.~(\ref{conf3}) can be reorganized as
\begin{align}
\mathcal K_i(E)f_i&=\kappa_i f_i,\qquad \kappa_2=-\kappa_1,\nonumber\\
\mathcal K_i(E)&=-\frac{d^2}{dz_i^2}-\frac1{z_i}\frac{d}{dz_i}
+a^2z_i^6+2ac_i z_i^4+
\left(\frac{b_1m}{\hbar}-\frac{2E}{\hbar^2}\right)z_i^2
+\frac{\Lambda^2}{\hbar^2z_i^2}.
\label{SturmSepM1}
\end{align}
Thus the physical energy is kept as a parameter of the operator, while the parabolic separation constant is diagonalized.

Set $\chi=z_i^2$ and use the same asymptotic factorization as in the Heun analysis,
\[
f_i(z_i)=\chi^s\exp\left(-\frac{a}{4}\chi^2-\frac{c_i}{2}\chi\right)P_i(\chi).
\]
The gauge-rotated form of Eq.~(\ref{SturmSepM1}) is
\begin{align}
\widetilde{\mathcal K}_i={}&-4\chi\frac{d^2}{d\chi^2}
+\left(4a\chi^2+4c_i\chi-8s-4\right)\frac{d}{d\chi}\nonumber\\
&+\left[4as+4a-c_i^2+\frac{b_1m}{\hbar}-\frac{2E}{\hbar^2}\right]\chi
+c_i(4s+2).
\label{KtM1add}
\end{align}
From Eq.~(\ref{KtM1add}), the polynomial space $\mathcal P_N=\langle1,\chi,\ldots,\chi^N\rangle$ is invariant when
\begin{eqnarray}
\frac{b_1m}{\hbar}-\frac{2E}{\hbar^2}=c_i^2-4a(N+s+1).
\label{QESM1add}
\end{eqnarray}
This is the same first termination condition as (\ref{HeunEnergyCondClean}), written in the Sturm parametrization.

On $\mathcal P_N$ we use the standard differential realization
\[
J^-=\frac{d}{d\chi},\qquad
J^0=\chi\frac{d}{d\chi}-\frac N2,\qquad
J^+=\chi^2\frac{d}{d\chi}-N\chi.
\]
Then
\begin{eqnarray}
\widetilde{\mathcal K}_{i,N}=-4J^0J^-+4aJ^++4c_iJ^0-(2N+8s+4)J^-+c_i(2N+4s+2),
\label{sl2M1add}
\end{eqnarray}
so Eq.~(\ref{sl2M1add}) shows explicitly that the separated Sturm operator belongs to $U(\mathfrak{sl}(2))$.  On the monomial basis its action is tridiagonal,
\[
\widetilde{\mathcal K}_{N}(c)\chi^j
=4a(j-N)\chi^{j+1}+c(4j+4s+2)\chi^j-4j(j+2s)\chi^{j-1}.
\]
If $Q_0=1$ and $Q_1=\kappa-c(4s+2)$, the characteristic polynomials therefore satisfy
\begin{align}
Q_{j+1}(\kappa)&=[\kappa-c(4j+4s+2)]Q_j(\kappa)
-16a\,j(j+2s)(N-j+1)Q_{j-1}(\kappa),\nonumber\\
&\hspace{5.6cm}1\leq j\leq N,\qquad Q_{N+1}(\kappa)=0.
\label{QrecM1add}
\end{align}
For illustration, $N=0$ gives $\kappa=c(4s+2)$ and $N=1$ gives
\[
\kappa_{\pm}=4c(s+1)\pm2\sqrt{c^2+4a(2s+1)}.
\]
Let $M_N(c)$ denote the matrix of $\widetilde{\mathcal K}_N(c)$ in the monomial basis. Because $c_2=-c_1$, conjugation by the diagonal involution $S\chi^j=(-1)^j\chi^j$ maps $M_N(c)$ to $-M_N(-c)$.  Thus the $\xi$- and $\eta$-separation spectra occur automatically in opposite pairs. 

Finally, (\ref{QESM1add}) gives
\begin{eqnarray}
E_{N,m}^{(\varepsilon)}=
\frac12\left[b_1\hbar m+2\hbar\Omega(N+1)+\varepsilon\Omega\Lambda-\frac{u_1^2}{\Omega^2}\right].
\label{EM1add}
\end{eqnarray}
Equation~(\ref{EM1add}) is exactly the positive-oscillator branch of the quadratic-algebra result (\ref{E1plusBranch}) after identifying $N=p$.  For the regular Frobenius exponent $\varepsilon=+1$, it also agrees with the cylindrical spectrum (\ref{M1cylEnergy}) after $N=n+\lambda$.  These explicit cross-checks show that the quadratic algebra, exact separation and the $\mathfrak{sl}(2)$ Sturm calculation describe the same finite energy sector, while Eq.~(\ref{QrecM1add}) determines the corresponding parabolic separation constants.

\section{The second superintegrable model}
We now turn to the second magnetic system of Ref.~\cite{Br1}.  We keep the two models separate because their magnetic fields and explicit integrals are different, but we follow the same calculational sequence as for Model I.  This makes it possible to see which features are model dependent and which belong to the common quadratic-algebra/Sturm mechanism.
The Hamiltonian is
\begin{eqnarray}
H=\frac{1}{2} (\mathbf{p^A})^2-\frac{b_1^2}{2}x_3^2 r^2+u_1 x_3+\frac{u_2}{r^2}+u_3(r^2+4x_3^2),\label{Hm1}
\end{eqnarray}
in the magnetic field
\begin{eqnarray}
\mathbf B=(b_1 x_1,b_1x_2,-2b_1 x_3)
\end{eqnarray}
with vector-potential components
\begin{eqnarray}
A_1=b_1 x_2 x_3,\quad A_2=-b_1x_1 x_3,\quad A_3=0.
\end{eqnarray}
Here $r^2=x_1^2+x_2^2$, while $x_3$ is the symmetry-axis coordinate.  As for Model I, we first determine the conserved operators and their polynomial closure before solving the separated equations.
The Hamiltonian (\ref{Hm1}) admits the following integrals of motion:
\begin{eqnarray}
&& X= \left(p_3^A\right)^2-2b_1 x_3 L_3^A+2u_1 x_3+8u_3 x_3^2-2b_1^2 x_3^2 r^2,
\\&&  Y= \frac{1}{2}(L_1^A p_2^A+p_2^A L_1^A)-\frac{1}{2}(L_2^A p_1^A+p_1^A L_2^A)-\frac{1}{2}b_1(r^2+4x_3^2)L_3^A+\frac{1}{2}u_1 r^2 - \frac{2u_2 x_3}{r^2}\nonumber\\&&\qquad +2u_3 x_3 r^2-\frac{1}{2}b_1^2 x_3 r^2(r^2+2x_3^2),
\\&& Z=L_3^A+b_1 x_3 r^2.
\end{eqnarray}

Superintegrability is reflected in the commutation relations
\begin{eqnarray}
[H,X]=0=[H,Y], \quad [H,Z]=[X,Z]=[Y,Z]=0.
\end{eqnarray}
As in the first model, $Z$ commutes with $H$, $X$ and $Y$ and therefore acts as a central element.  The corresponding diagram is

\begin{eqnarray}
\begin{xy}
(0,0)*+{X}="m"; (20,0)*+{Z}="q"; (40,0)*+{Y}="p"; (20,20)*+{H}="r"; 
"r";"p"**\dir{--};
"m";"q"**\dir{--}; % corrected: commuting edge X--Z (was X--Y)
"q";"p"**\dir{--}; 
"r";"m"**\dir{--};
"r";"q"**\dir{--};
\end{xy}
\end{eqnarray}
A direct computation of the remaining commutators gives the quadratic algebra
\begin{eqnarray}
&&[X,Y]=W,\nonumber
\\&& [X,W]=4\hbar^2(u_1-b_1 Z)X+32\hbar^2 u_3 Y-8\hbar^2(u_1-b_1 Z)H,\label{QA2}
\\&& [Y,W]=6\hbar^2 X^2-16\hbar^2 H X-4\hbar^2(u_1-b_1 Z)Y+ 8\hbar^2 H^2-16\hbar^2 u_3 Z^2-16\hbar^2u_3(2u_2-\hbar^2).\nonumber
\end{eqnarray}
This quadratic algebra can be written as 
\begin{eqnarray}
&&[X,Y]=W, \quad [X,W]= \delta X + \epsilon Y + \zeta, \nonumber
\\&& [Y,W]= a X^2 + d X - \delta Y + z \nonumber
\end{eqnarray}
by setting 
\begin{eqnarray}
&&\delta=4\hbar^2 (u_1-b_1 Z),\quad \epsilon=32\hbar^2 u_3,\quad \zeta= -8\hbar^2 (u_1 -b_1 Z)H,\nonumber\\&& a = 6\hbar^2,\quad d=-16\hbar^2 H,\quad z = 8\hbar^2 H^2 - 16\hbar^2 u_3 Z^2  - 16\hbar^2 u_3(2u_2- \hbar^2). \nonumber
\end{eqnarray}
The corresponding abstract Casimir is
\begin{align}
K={}&W^2-4\hbar^2\mathcal U\{X,Y\}-32\hbar^2u_3 Y^2
+16\hbar^2\mathcal U H Y+4\hbar^2 X^3-16\hbar^2 H X^2\nonumber\\
&+16\hbar^2\left[H^2-2u_3Z^2+2u_3(3\hbar^2-2u_2)\right]X,
\qquad \mathcal U=u_1-b_1 Z.
\label{KabsM2add}
\end{align}
Substitution of the differential realization reduces the same Casimir to an expression involving only the central elements,
\begin{eqnarray}
&& K= 4 b_1^2 \hbar^2 Z^4-8 b_1 u_1 \hbar^2 Z^3+128 u_3\hbar^4 H -4\hbar^2 (b_1^2 \hbar^2 -  u_1^2 - 2 b_1^2 u_2)Z^2 \nonumber\\&&\qquad + 8 \hbar^2 (b_1 u_1 \hbar^2 - 2 b_1 u_1 u_2)Z -4\hbar^2 ( u_1^2 \hbar^2 - 2 u_1^2 u_2).
\end{eqnarray}

\subsection{Structure function and energy spectrum}
The representation analysis now proceeds in direct parallel with Sect.~3.  The important simplification is that the combination $u_1-b_1Z$ is central and therefore behaves as an effective linear coupling inside each irreducible sector.  On a $Z=m\hbar$ eigenspace introduce
\begin{eqnarray}
\mathcal U_m=u_1-b_1m\hbar,\qquad
\Omega_2^2=8u_3,\qquad
\Lambda^2=m^2\hbar^2+2u_2.
\label{M2scalesadd}
\end{eqnarray}
Here $\epsilon=4\hbar^2\Omega_2^2$, so the deformed-oscillator realization can be displayed explicitly rather than passing directly from the quadratic algebra to its structure function.  On the $(E,m)$ eigenspace one obtains
\begin{align}
X={}&2\hbar\Omega_2(\mathcal N+q),\nonumber\\
Y={}&-\frac{2\hbar\mathcal U_m}{\Omega_2}(\mathcal N+q)
+\frac{2\mathcal U_m E}{\Omega_2^2}
+b^\dagger\rho_2(\mathcal N)+\rho_2(\mathcal N)b,\nonumber\\
W={}&2\hbar\Omega_2[b^\dagger\rho_2(\mathcal N)-\rho_2(\mathcal N)b],
\label{M2oscReal}
\end{align}
where
\[
\rho_2(\mathcal N)=\left(\mathcal N+q+1-\eta_2\right)^{-1/2},
\qquad
\eta_2=\frac{\hbar\Omega_2^3-\mathcal U_m^2}{2\hbar\Omega_2^3}.
\]
As for Model I, this choice of $\rho_2$ is a normalization of the oscillator ladders.  The canonical choice $\rho_2=1$ gives the equivalent cubic structure function. The present choice produces the quartic factorization used for the endpoint analysis.  Substitution of (\ref{M2oscReal}) into the remaining quadratic-algebra relation and the Casimir gives 
\begin{align}
\phi_2(t,E,q)={}&2\hbar^3\Omega_2
\left(t+q-\frac{\hbar\Omega_2^3-\mathcal U_m^2}{2\hbar\Omega_2^3}\right)^2\nonumber\\
&\times\left(t+q-\frac{2E+\hbar\Omega_2-\Omega_2\Lambda}{2\hbar\Omega_2}\right)
\left(t+q-\frac{2E+\hbar\Omega_2+\Omega_2\Lambda}{2\hbar\Omega_2}\right).
\label{phiM2add}
\end{align}
Here $t$ denotes the eigenvalue variable associated with the number operator. $\phi_2(t,E,q)$ is viewed as a polynomial in $t$ and it contains the information on the spectrum. Imposing $\phi_2(0)=\phi_2(p+1)=0$ gives the branch associated with the decaying oscillator gauge,
\begin{eqnarray}
E_{2,+}^{(\varepsilon)}=
\frac12\left[2\hbar\Omega_2(p+1)+\varepsilon\Omega_2\Lambda-\frac{\mathcal U_m^2}{\Omega_2^2}\right],\qquad \varepsilon=\pm1.
\label{E2plusadd}
\end{eqnarray}
The opposite formal algebraic branch is
\begin{eqnarray}
E_{2,-}^{(\varepsilon)}=
\frac12\left[-2\hbar\Omega_2(p+1)+\varepsilon\Omega_2\Lambda-\frac{\mathcal U_m^2}{\Omega_2^2}\right].
\end{eqnarray}
For the first branch, $q=\eta_2$, and hence $\rho_2^2(\mathcal N)=1/(\mathcal N+1)>0$.  Thus the standard positivity condition applies directly to the corresponding structure function.  For $\Omega_2>0$, the regular choice $\varepsilon=+1$ gives the physical branch, which will be recovered from the separated Schr\"{o}dinger equation below.  The opposite branch is retained as an additional algebraic solution of the endpoint conditions but is not used below to identify the regular physical spectrum.

\section{Separation of variables for the second model}
We now repeat the same three-step analysis for Model II: parabolic separation, cylindrical separation, and the finite $\mathfrak{sl}(2)$ Sturm algebraization of the parabolic equations.  Keeping the organization parallel to Sect.~4 makes it easier to distinguish genuine differences in the parameters from features that are common to both magnetic systems.
\subsection{Parabolic coordinates}
The Schr\"{o}dinger equation $H\Psi(\xi,\eta,\phi)=E\Psi(\xi,\eta,\phi)$ for (\ref{Hm1}), written in the circular parabolic coordinates introduced in Sect.~4, is
\begin{align}
0={}&-\frac{\hbar}{\xi^2+\eta^2}\Bigg[
\hbar\xi^2\eta^2\frac{\partial^2\Psi}{\partial\xi^2}
+\hbar\xi\eta^2\frac{\partial\Psi}{\partial\xi}
+\hbar\xi^2\eta^2\frac{\partial^2\Psi}{\partial\eta^2}
+\hbar\xi^2\eta\frac{\partial\Psi}{\partial\eta}\nonumber\\
&\qquad+\hbar(\xi^2+\eta^2)\frac{\partial^2\Psi}{\partial\phi^2}
+i b_1\xi^2\eta^2(\eta^4-\xi^4)\frac{\partial\Psi}{\partial\phi}\Bigg]\nonumber\\
&+\Big[2u_3\xi^2\eta^2(\xi^4-\xi^2\eta^2+\eta^4)
+u_1\xi^2\eta^2(\xi^2-\eta^2)-2E\xi^2\eta^2+2u_2\Big]\Psi.
\label{Schr2}
\end{align}
Using the same circular parabolic coordinates as in Sect.~4 and substituting $\Psi(\xi,\eta,\phi)=f_1(\xi)f_2(\eta)e^{im\phi}$ into Eq.~(\ref{Schr2}), the equation separates into
\begin{align}
0={}&\frac{\partial^2 f_1(\xi)}{\partial \xi^2}+\frac{1}{\xi}\frac{\partial f_1(\xi)}{\partial \xi}\nonumber\\
&-\left(2u'_3 \xi^6+(u'_1-b'_1 m)\xi^4-2E'\xi^2+\frac{2u'_2+m^2}{\xi^2}+k'_1\right)f_1(\xi),\label{eqq1}\\
0={}&\frac{\partial^2 f_2(\eta)}{\partial \eta^2}+\frac{1}{\eta}\frac{\partial f_2(\eta)}{\partial \eta}\nonumber\\
&-\left(2u'_3 \eta^6-(u'_1-b'_1 m)\eta^4-2E'\eta^2+\frac{2u'_2+m^2}{\eta^2}-k'_1\right)f_2(\eta).\label{eqq2}
\end{align}
Here $u'_1=u_1/\hbar^2$, $u'_2=u_2/\hbar^2$, $u'_3=u_3/\hbar^2$, $b'_1=b_1/\hbar$, $E'=E/\hbar^2$ and $k'_1=k_1/\hbar^2$.
With $z_1=\xi$, $z_2=\eta$ the two equations can again be written in the common form
\begin{equation}
\frac{d^2 f_i(z_i)}{d z_i^2}+\frac{1}{z_i}\frac{d f_i(z_i)}{d z_i}
-\left(2u'_3 z_i^6+\tau_i z_i^4-2E'z_i^2+\frac{\sigma}{z_i^2}+k'_i\right)f_i(z_i)=0.
\label{M2CommonParabolic}
\end{equation}
Here $i=1,2$, $k'_2=-k'_1$, $\tau_1=u'_1-b'_1m=-\tau_2$ and $\sigma=m^2+2u'_2$.  Thus Model II leads to the same sextic singular-oscillator structure as Model I, with a different parameter dependence in the quartic and quadratic coefficients.  The local behavior is again governed by the inverse-square term, while the sextic term fixes the decaying gauge factor.  Consequently the same change $\chi=z_i^2$ and the same type of exponential/Frobenius factor reduce the parabolic equations to biconfluent-Heun form.  Rather than repeat the full parameter matching of Sect.~4.1, we give the equivalent finite $\mathfrak{sl}(2)$ construction explicitly in Sect.~6.3.

\subsection{Cylindrical coordinates}
Cylindrical coordinates again provide the most direct route to the normalizable bound-state wavefunctions.  In these coordinates the Schr\"{o}dinger equation $H\Psi(\rho,\phi,z)=E\Psi(\rho,\phi,z)$ reads
\begin{align}
\Bigg[&-\frac{\hbar^2}{2}\left(\frac{\partial^2}{\partial\rho^2}
+\frac1\rho\frac{\partial}{\partial\rho}
+\frac1{\rho^2}\frac{\partial^2}{\partial\phi^2}
+\frac{\partial^2}{\partial z^2}\right)
+i\hbar b_1z\frac{\partial}{\partial\phi}
+u_1z+\frac{u_2}{\rho^2}+u_3(\rho^2+4z^2)\Bigg]\Psi=E\Psi.
\label{Cylin2}
\end{align}
Here again the scalar term cancels the $A^2/2$ contribution and after $\Psi\propto e^{im\phi}$ the effective linear coupling in $z$ is $u_1-b_1m\hbar$.
 
Using $\Psi(\rho,\phi,z)=f_1(\rho)f_2(z)e^{im\phi}$, Eq.~(\ref{Cylin2}) separates into
\begin{align}
\frac{d^2f_1}{d\rho^2}+\frac{1}{\rho}\frac{df_1}{d\rho}
+\left(2E'-k_1-2u'_3\rho^2-\frac{m^2+2u'_2}{\rho^2}\right)f_1&=0,\nonumber\\
\frac{d^2f_2}{dz^2}+\left[k_1+2(b'_1m-u'_1)z-8u'_3z^2\right]f_2&=0.
\label{M2CylODEs}
\end{align}
Here $u'_1=\frac{u_1}{\hbar^2}$, $u'_2=\frac{u_2}{\hbar^2}$, $u'_3=\frac{u_3}{\hbar^2}$, $b'_1=\frac{b_1}{\hbar}$, $E'=\frac{E}{\hbar^2}$, and $k_1$ is the separation constant.

The two equations in (\ref{M2CylODEs}) are exactly solvable, providing an independent check of the algebraic construction.  Using the scales in Eq.~(\ref{M2scalesadd}), write $a_2=\Omega_2/(2\hbar)$, $d_2=\mathcal U_m/\hbar^2$, and $\sigma_2=\Lambda^2/\hbar^2$.  The normalizable solutions are of Laguerre and Hermite type,
\begin{align*}
R_{n_r}(\rho)&=\rho^{\sqrt{\sigma_2}}e^{-a_2\rho^2/2}
L_{n_r}^{(\sqrt{\sigma_2})}(a_2\rho^2),\\
Z_{n_z}(z)&=e^{-a_2\left(z+\frac{d_2}{4a_2^2}\right)^2}
H_{n_z}\left(\sqrt{2a_2}\left(z+\frac{d_2}{4a_2^2}\right)\right).
\end{align*}
Here $L_{n_r}^{(\sqrt{\sigma_2})}$ is a generalized Laguerre polynomial and $H_{n_z}$ is a Hermite polynomial.  The integers $n_r,n_z=0,1,2,\ldots$ count the radial and longitudinal excitations, respectively.
Combining the two quantization conditions gives, with the total excitation number $N=n_r+n_z$,
\begin{eqnarray}
E=2\hbar\sqrt{2u_3}(N+1)+\sqrt{2u_3}\,\Lambda
-\frac{(u_1-b_1m\hbar)^2}{16u_3}.
\label{M2cylEadd}
\end{eqnarray}
The regular choice corresponds to the positive Frobenius exponent at $\rho=0$; the second local exponent is obtained formally by $\Lambda\mapsto-\Lambda$ and is physical only when the boundary conditions allow it.

\subsection{Sturm form and explicit \texorpdfstring{$\mathfrak{sl}(2)$}{sl(2)} algebraization}
The parabolic equation (\ref{M2CommonParabolic}) also admits a one-variable Sturm algebraization.  We give the operator explicitly, rather than referring back to Model I, so that the parallel structure is visible:
\begin{align}
\mathcal K_i^{(2)}(E)f_i&=\kappa_i f_i,\qquad \kappa_2=-\kappa_1,\nonumber\\
\mathcal K_i^{(2)}(E)&=-\frac{d^2}{dz_i^2}-\frac1{z_i}\frac{d}{dz_i}
+2u'_3z_i^6+\tau_i z_i^4-2E'z_i^2+\frac{\sigma}{z_i^2}.
\label{M2SturmSepExplicit}
\end{align}
Here $\kappa_i=-k_i^\prime$, $\tau_1=(u_1-b_1m\hbar)/\hbar^2=-\tau_2$ and $\sigma=(m^2\hbar^2+2u_2)/\hbar^2$.  As in Model I, $E$ remains a parameter of the Sturm operator while the separation constant is diagonalized.  Set $a_2=\Omega_2/(2\hbar)$, $c_1=\mathcal U_m/(\hbar\Omega_2)$, $c_2=-c_1$, and $s=\varepsilon\Lambda/(2\hbar)$.
With $\chi=z_i^2$, the gauge transformation
\[
f_i(z_i)=\chi^s\exp\left(-\frac{a_2}{4}\chi^2-\frac{c_i}{2}\chi\right)P_i(\chi)
\]
gives
\begin{align}
\widetilde{\mathcal K}^{(2)}_i={}&-4\chi\frac{d^2}{d\chi^2}
+\left(4a_2\chi^2+4c_i\chi-8s-4\right)\frac{d}{d\chi}\nonumber\\
&+\left(4a_2s+4a_2-c_i^2-\frac{2E}{\hbar^2}\right)\chi+c_i(4s+2).
\label{M2sl2opadd}
\end{align}
From Eq.~(\ref{M2sl2opadd}), the coefficient of the raising term vanishes on $\mathcal P_N$ when $-2E/\hbar^2=c_i^2-4a_2(N+s+1)$.  This invariant-space condition yields
\begin{eqnarray}
E_{N,m}^{(2,\varepsilon)}=
\frac12\left[2\hbar\Omega_2(N+1)+\varepsilon\Omega_2\Lambda-
\frac{\mathcal U_m^2}{\Omega_2^2}\right].
\label{M2QESadd}
\end{eqnarray}
On $\mathcal P_N=\langle1,\chi,\ldots,\chi^N\rangle$ use the standard generators
\[
J^-_{(2)}=\frac{d}{d\chi},\qquad
J^0_{(2)}=\chi\frac{d}{d\chi}-\frac N2,\qquad
J^+_{(2)}=\chi^2\frac{d}{d\chi}-N\chi.
\]
Then the second-model Sturm operator takes the explicit enveloping-algebra form
\begin{eqnarray}
\widetilde{\mathcal K}^{(2)}_{i,N}=-4J^0_{(2)}J^-_{(2)}+4a_2J^+_{(2)}+4c_iJ^0_{(2)}
-(2N+8s+4)J^-_{(2)}+c_i(2N+4s+2).
\label{M2sl2Explicit}
\end{eqnarray}
Equation~(\ref{M2sl2Explicit}) shows that each separated Sturm equation belongs to $U(\mathfrak{sl}(2))$ on $\mathcal P_N$.  The monomial action is tridiagonal,
\[
\widetilde{\mathcal K}^{(2)}_N(c_i)\chi^j
=4a_2(j-N)\chi^{j+1}+c_i(4j+4s+2)\chi^j-4j(j+2s)\chi^{j-1}.
\]
Thus, with $Q^{(2)}_0=1$ and $Q^{(2)}_1=\kappa-c_i(4s+2)$,
\begin{align}
Q^{(2)}_{j+1}(\kappa)&=[\kappa-c_i(4j+4s+2)]Q^{(2)}_j(\kappa)
-16a_2j(j+2s)(N-j+1)Q^{(2)}_{j-1}(\kappa),\nonumber\\
&\hspace{5.1cm}1\leq j\leq N,\qquad Q^{(2)}_{N+1}(\kappa)=0.
\label{M2QrecExplicit}
\end{align}
For illustration, $N=0$ gives $\kappa=c_i(4s+2)$ and $N=1$ gives
\[
\kappa_{\pm}=4c_i(s+1)\pm2\sqrt{c_i^2+4a_2(2s+1)}.
\]
Let $M_N^{(2)}(c)$ denote the matrix of $\widetilde{\mathcal K}^{(2)}_N(c)$ in the monomial basis. Since $c_2=-c_1$, the diagonal involution $S\chi^j=(-1)^j\chi^j$ maps $M_N^{(2)}(c)$ to $-M_N^{(2)}(-c)$, so the $\xi$- and $\eta$-separation spectra again occur in opposite pairs without the need for a separate numbered formula.  Finally, Eq.~(\ref{M2QESadd}) agrees with the quadratic-algebra spectrum (\ref{E2plusadd}) for $N=p$ and, for the regular positive exponent, with the cylindrical result (\ref{M2cylEadd}).  Equation~(\ref{M2QrecExplicit}) supplies the corresponding parabolic separation constants.

\section{Sturm representation of the complete Hamiltonians}\label{fullSturmAdded}
The $\mathfrak{sl}(2)$ constructions of Sects.~4 and 6 start only after parabolic separation.  We now ask whether the complete two-variable problem also has a useful algebraic form before separation.  The answer is affirmative for both models.  The inverse-square term is particularly convenient for this purpose: after fixing the axial integral, multiplication of the Schr\"{o}dinger equation by $\rho^2$ converts the coupling $u_2/\rho^2$ into an additive spectral term.  We therefore keep the physical energy $E$ fixed and treat $u_2$ as the Sturm eigenvalue, transform the commuting integrals at the operator level, and only then gauge rotate the resulting two-variable problem.  This follows the same general logic as the Sturm representation used in Ref.~\cite{Quesne2026}: the spectral role of a coupling is exchanged with that of the energy parameter, while the invariance operators are transformed together with the Hamiltonian.  The specific Lie algebra obtained here is the projective $\mathfrak{sl}(3)$ algebra appropriate to polynomials in two variables.

Before treating the two models separately, we record the projective $\mathfrak{sl}(3)$ realization used for both.  Its advantage is that all generators map polynomials of total degree at most $N$ back into the same finite triangular space.  The required first-order operators are
\begin{eqnarray}
J_1^-&=&\partial_x,\qquad J_2^-=\partial_y,\nonumber\\
J_{11}&=&x\partial_x,\quad J_{12}=x\partial_y,\quad
J_{21}=y\partial_x,\quad J_{22}=y\partial_y,\nonumber\\
J_1^+&=&x(x\partial_x+y\partial_y-N),\qquad
J_2^+=y(x\partial_x+y\partial_y-N).
\label{sl3gensAdded}
\end{eqnarray}
These generators preserve the triangular polynomial space \cite{Turbiner1992}
\begin{eqnarray}
\mathcal P_N^{(2)}=\langle x^iy^j\,|\,i,j\ge0,\ i+j\le N\rangle,
\qquad \dim\mathcal P_N^{(2)}=\frac{(N+1)(N+2)}2.
\label{PNsl3Added}
\end{eqnarray}
The space $\mathcal P_N^{(2)}$ is the finite invariant module of the gauge-rotated differential operator.  At a fixed Sturm eigenvalue, the physical solution space is the appropriate eigenspace selected inside this triangular module; in the separated oscillator basis below, the level $p=N$ has multiplicity $N+1$ rather than the full triangular dimension.

\subsection{First magnetic model: full \texorpdfstring{$\mathfrak{sl}(3)$}{sl(3)} algebraization}
We begin with Model I.  The goal is not to replace the separated $\mathfrak{sl}(2)$ description developed in Sect.~4, but to identify the algebra that acts naturally on the complete two-variable Sturm problem from which those separated equations originate.  For the first model the central integral $Z$ reduces to the canonical $L_z$.  On $Z=m\hbar$ define
\begin{eqnarray}
\omega_1^2=2(4u_3+b_2m\hbar),\qquad g_1=u_1,\qquad \Delta_1=\frac12b_1m\hbar.
\label{M1fullparsAdd}
\end{eqnarray}
Because $Z=L_z$ in this gauge, restricting to $Z=m\hbar$ removes the azimuthal variable and gives the two-variable Hamiltonian
\begin{align}
H_m^{(1)}={}&-\frac{\hbar^2}{2}\left(\partial_\rho^2+\frac1\rho\partial_\rho+\partial_z^2\right)
+\frac{m^2\hbar^2+2u_2}{2\rho^2}
+\frac{\omega_1^2}{8}\rho^2+\frac{\omega_1^2}{2}z^2+g_1z+\Delta_1.
\label{H1fixedAdd}
\end{align}
The coupling $u_2$ occurs only in the centrifugal term and is therefore a natural Sturm parameter.  Removing it from the operator and moving it to the right-hand side gives
\begin{eqnarray}
\mathscr K_1(E)=\rho^2\left(H_m^{(1)}\big|_{u_2=0}-E\right),\qquad
\mathscr K_1(E)\Psi=-u_2\Psi.
\label{K1fullAdd}
\end{eqnarray}
The integral $X$ does not contain $u_2$.  If
\begin{eqnarray}
Y=Y_0^{(1)}-\frac{2u_2z}{\rho^2},
\end{eqnarray}
then the corresponding Sturm integral is
\begin{eqnarray}
\mathscr Y_1=Y_0^{(1)}+\frac{2z}{\rho^2}\mathscr K_1.
\label{B1fullAdd}
\end{eqnarray}
A direct calculation then gives
\begin{eqnarray}
[\mathscr K_1,X]=[\mathscr K_1,\mathscr Y_1]=0.
\end{eqnarray}
Thus the Sturm transformation preserves the two nontrivial symmetries: $X$ is unchanged because it contains no $u_2$, whereas $Y$ acquires the compensating term in (\ref{B1fullAdd}).
With $\mathscr W_1=[X,\mathscr Y_1]$, and with $Z$ kept as a central operator, put
\begin{eqnarray}
\Omega_1^2=2(4u_3+b_2Z),\qquad Z_1=\frac12b_1Z.
\end{eqnarray}
The transformed integrals close as
\begin{eqnarray}
[X,\mathscr Y_1]&=&\mathscr W_1,\nonumber\\
{}[X,\mathscr W_1]&=&4\hbar^2u_1 X+4\hbar^2\Omega_1^2\mathscr Y_1
-8\hbar^2u_1(E-Z_1),\label{M1SturmAlgAdd}\\
{}[\mathscr Y_1,\mathscr W_1]&=&6\hbar^2 X^2-16\hbar^2(E-Z_1)X-4\hbar^2u_1\mathscr Y_1
+8\hbar^2(E-Z_1)^2\nonumber\\
&&+2\hbar^2\Omega_1^2(\hbar^2-Z^2+2\mathscr K_1).
\end{eqnarray}
Thus $\mathscr K_1$ itself is a central element of the Sturm symmetry algebra.

To algebraize the complete differential operators, set
\begin{eqnarray}
x=\rho^2,\qquad y=z+\frac{g_1}{\omega_1^2},\qquad
\overline E_1=E-\Delta_1+\frac{g_1^2}{2\omega_1^2}.
\end{eqnarray}
The shift in $y$ completes the square in the longitudinal oscillator and removes the linear $g_1z$ term.  We then factor out the Gaussian and Frobenius behavior with the gauge factor
\begin{eqnarray}
\Gamma_r^{(1)}=x^r\exp\left[-\frac{\omega_1x}{4\hbar}-\frac{\omega_1y^2}{2\hbar}\right].
\end{eqnarray}
The exponent $r$ is the Frobenius exponent associated with the inverse-square term.  On a physical Sturm eigenspace it is related to the coupling by
\begin{eqnarray}
2\hbar r=\Lambda=\sqrt{m^2\hbar^2+2u_2},\qquad
u_2=\frac{\hbar^2}{2}(4r^2-m^2).
\label{rCouplingM1clean}
\end{eqnarray}
In the Sturm problem $r$ is determined algebraically rather than fixed in advance.  The polynomial space (\ref{PNsl3Added}) is invariant when
\begin{eqnarray}
\overline E_1=\hbar\omega_1(N+r+1).
\label{M1sl3condAdd}
\end{eqnarray}
Under this invariant-space condition, after the gauge rotation, the Sturm Hamiltonian is
\begin{eqnarray}
\widetilde{\mathscr K}_1=
-2\hbar^2(J_{11}^2+2rJ_{11})-\frac{\hbar^2}{2}J_{12}J_2^-
+\hbar\omega_1J_1^++\hbar^2\left(\frac{m^2}{2}-2r^2\right).
\label{K1sl3Add}
\end{eqnarray}
The two commuting integrals are algebraized by the same generators,
\begin{eqnarray}
\widetilde X=-\hbar^2(J_2^-)^2+2\hbar\omega_1J_{22}
+\hbar\omega_1-\frac{g_1^2}{\omega_1^2},
\label{A1sl3Add}
\end{eqnarray}
\begin{equation}
\begin{aligned}
\widetilde{\mathscr Y}_1={}&-2\hbar^2J_{11}J_2^-
-\hbar^2J_{22}J_2^-+\frac{\hbar^2g_1}{\omega_1^2}(J_2^-)^2
+2\hbar\omega_1J_2^++\frac{\hbar\omega_1}{2}J_{12}\\
&-\hbar^2(2r+1)J_2^--\frac{2\hbar g_1}{\omega_1}J_{22}
+\frac{\hbar g_1}{\omega_1}(2N+2r+1).
\end{aligned}
\label{B1sl3Add}
\end{equation}
Equations (\ref{K1sl3Add})--(\ref{B1sl3Add}) show that the Sturm Hamiltonian and both non-central integrals are represented in the same $U(\mathfrak{sl}(3))$.  This is stronger than algebraizing the Hamiltonian alone: the complete symmetry problem is carried into one common finite-dimensional polynomial representation.

The Sturm problem is also explicitly solvable.  Define
\begin{eqnarray}
\Lambda_{1,p}(E)=\frac{2}{\omega_1}[\overline E_1-\hbar\omega_1(p+1)].
\end{eqnarray}
Then
\begin{eqnarray}
u_{2,p}^{(1)}(E)=\frac12[\Lambda_{1,p}(E)^2-m^2\hbar^2]
=\frac{2}{\omega_1^2}[\overline E_1-\hbar\omega_1(p+1)]^2-\frac{m^2\hbar^2}{2}.
\label{M1SturmSpecAdd}
\end{eqnarray}
For $\Lambda_{1,p}\ge0$ the corresponding $p+1$ states are
\begin{align}
\Psi_{p;n_\rho,n_z}^{(1)}={}&e^{im\phi}\rho^{\Lambda_{1,p}/\hbar}
\exp\left[-\frac{\omega_1\rho^2}{4\hbar}
-\frac{\omega_1}{2\hbar}\left(z+\frac{g_1}{\omega_1^2}\right)^2\right]\nonumber\\
&\times L_{n_\rho}^{(\Lambda_{1,p}/\hbar)}\left(\frac{\omega_1\rho^2}{2\hbar}\right)
H_{n_z}\left(\sqrt{\frac{\omega_1}{\hbar}}\left(z+\frac{g_1}{\omega_1^2}\right)\right),
\qquad n_\rho+n_z=p.
\label{M1SturmWFAdd}
\end{align}
Thus the full Sturm problem gives both the allowed values of the coupling and their wavefunctions.  For a fixed polynomial degree $p$ the oscillator quantum numbers satisfy $n_\rho+n_z=p$, giving the expected $p+1$ states at that level.  Parabolic separation of the same Sturm eigenspace leads to the finite $(p+1)$-dimensional $\mathfrak{sl}(2)$ problem of Sect.~4.  The $\mathfrak{sl}(3)$ and $\mathfrak{sl}(2)$ descriptions therefore refer to the unreduced and separated versions of the same Sturm problem.

\subsection{Second magnetic model: full \texorpdfstring{$\mathfrak{sl}(3)$}{sl(3)} algebraization}
We now repeat the operator-level construction for Model II.  The steps are deliberately displayed rather than inferred from Model I, because this makes clear that the second system has the same $\mathfrak{sl}(3)$-before-separation/$\mathfrak{sl}(2)$-after-separation hierarchy.  For the second model $Z$ again reduces to $L_z$.  On $Z=m\hbar$ put
\begin{eqnarray}
\omega_2^2=8u_3,\qquad g_{2,m}=u_1-b_1m\hbar.
\label{M2fullparsAdd}
\end{eqnarray}
The Hamiltonian becomes
\begin{align}
H_m^{(2)}={}&-\frac{\hbar^2}{2}\left(\partial_\rho^2+\frac1\rho\partial_\rho+\partial_z^2\right)
+\frac{m^2\hbar^2+2u_2}{2\rho^2}
+\frac{\omega_2^2}{8}\rho^2+\frac{\omega_2^2}{2}z^2+g_{2,m}z.
\label{H2fixedAdd}
\end{align}
At this point it is useful to clarify why the same $\mathfrak{sl}(3)$ algebraization appears for two systems that are distinct in the original magnetic classification.  The classification in Ref.~\cite{Br1} concerns the complete three-dimensional magnetic Hamiltonians: their magnetic fields, vector potentials and explicit integrals are different, and no gauge, canonical or parameter transformation between the two original systems is implied here.  What becomes common only after symmetry reduction is the functional form of the two-variable scalar problem.  Indeed, after restriction to an axial sector $Z=m\hbar$, Eqs.~(\ref{H1fixedAdd}) and (\ref{H2fixedAdd}) are both specializations of
\begin{align}
H_{\rm red}(\omega,g,\Delta)={}&-\frac{\hbar^2}{2}\left(\partial_\rho^2+\frac1\rho\partial_\rho+\partial_z^2\right)
+\frac{m^2\hbar^2+2u_2}{2\rho^2}
+\frac{\omega^2}{8}\rho^2+\frac{\omega^2}{2}z^2+gz+\Delta.
\label{commonReducedForm}
\end{align}
The different magnetic embeddings of the effective parameters are summarized by
\begin{center}
\begin{tabular}{c|c|c|c}
 & $\omega^2$ & $g$ & $\Delta$ \\ \hline
Model I & $2(4u_3+b_2Z)$ & $u_1$ & $\frac12 b_1Z$ \\
Model II & $8u_3$ & $u_1-b_1Z$ & $0$
\end{tabular}
\end{center}
with $Z=m\hbar$ on the reduced sector.  This table also shows what remains model dependent.  In Model I the axial quantum number enters the effective oscillator frequency and the additive shift, whereas in Model II the oscillator frequency is independent of $m$ and the axial quantum number instead enters the linear coupling.  Consequently, even though every fixed-$m$ sector belongs to the same reduced singular $2{:}1$ oscillator family, the way these sectors are assembled into the full three-dimensional spectrum is different in the two magnetic models.  The common projective $\mathfrak{sl}(3)$ realization is therefore a property of the reduced/Sturm normal form (\ref{commonReducedForm}), not evidence that the original magnetic systems are equivalent.

As for Model I, the inverse-square coupling is isolated by defining
\begin{eqnarray}
\mathscr K_2(E)=\rho^2\left(H_m^{(2)}\big|_{u_2=0}-E\right),\qquad
\mathscr K_2(E)\Psi=-u_2\Psi.
\label{K2fullAdd}
\end{eqnarray}
Writing $Y=Y_0^{(2)}-2u_2z/\rho^2$, the transformed integral
\begin{eqnarray}
\mathscr Y_2=Y_0^{(2)}+\frac{2z}{\rho^2}\mathscr K_2.
\label{B2fullAdd}
\end{eqnarray}
satisfies
\begin{eqnarray}
[\mathscr K_2,X]=[\mathscr K_2,\mathscr Y_2]=0.
\end{eqnarray}
Let $G_2=u_1-b_1Z$ and $\mathscr W_2=[X,\mathscr Y_2]$.  The transformed quadratic algebra is
\begin{eqnarray}
[X,\mathscr Y_2]&=&\mathscr W_2,\nonumber\\
{}[X,\mathscr W_2]&=&4\hbar^2G_2 X+32\hbar^2u_3\mathscr Y_2-8\hbar^2G_2 E,\\
{}[\mathscr Y_2,\mathscr W_2]&=&6\hbar^2 X^2 -16\hbar^2 E X-4\hbar^2G_2\mathscr Y_2+8\hbar^2 E^2
+16\hbar^2u_3(\hbar^2-Z^2+2\mathscr K_2).
\label{M2SturmAlgAdd}
\end{eqnarray}

The common reduction can also be seen directly at the level of the transformed quadratic algebra.  Introduce the reduced energy $\mathcal E=E-\Delta$ and use the effective parameters $(\omega,g,\Delta)$ of Eq.~(\ref{commonReducedForm}).  Both transformed quadratic algebras then take the single form
\begin{align}
[X,\mathscr Y]&=\mathscr W,\nonumber\\
[X,\mathscr W]&=4\hbar^2gX+4\hbar^2\omega^2\mathscr Y-8\hbar^2g\mathcal E,\nonumber\\
[\mathscr Y,\mathscr W]&=6\hbar^2X^2-16\hbar^2\mathcal E X-4\hbar^2g\mathscr Y
+8\hbar^2\mathcal E^2
+2\hbar^2\omega^2\bigl(\hbar^2-Z^2+2\mathscr K\bigr).
\label{commonSturmAlgebra}
\end{align}
For Model I one substitutes $(\omega^2,g,\Delta,\mathscr K)=(\Omega_1^2,u_1,\frac12b_1Z,\mathscr K_1)$, whereas for Model II one uses $(8u_3,u_1-b_1Z,0,\mathscr K_2)$.  Thus the coincidence is already present before the explicit $\mathfrak{sl}(3)$ realization: after axial reduction and the Sturm transformation, the two systems possess the same quadratic-algebra normal form with different effective parameters.

As in Model I, we first shift the longitudinal coordinate so that the oscillator is centered and then use the same radial variable.  Set
\begin{eqnarray}
x=\rho^2,\qquad y=z+\frac{g_{2,m}}{\omega_2^2},\qquad
\overline E_2=E+\frac{g_{2,m}^2}{2\omega_2^2}.
\end{eqnarray}
The shifted coordinate removes the linear term $g_{2,m}z$.  We factor out the same Gaussian and Frobenius behavior with the gauge factor
\begin{eqnarray}
\Gamma_r^{(2)}=x^r\exp\left[-\frac{\omega_2x}{4\hbar}-\frac{\omega_2y^2}{2\hbar}\right].
\end{eqnarray}
The same radial relation holds,
\begin{eqnarray}
2\hbar r=\Lambda=\sqrt{m^2\hbar^2+2u_2},\qquad
u_2=\frac{\hbar^2}{2}(4r^2-m^2).
\end{eqnarray}
Thus the Sturm eigenvalue fixes the allowed Frobenius exponent.  The finite polynomial module is obtained when
\begin{eqnarray}
\overline E_2=\hbar\omega_2(N+r+1).
\label{M2sl3condAdd}
\end{eqnarray}
Under this condition, the gauge-rotated Sturm Hamiltonian is
\begin{eqnarray}
\widetilde{\mathscr K}_2=
-2\hbar^2(J_{11}^2+2rJ_{11})-\frac{\hbar^2}{2}J_{12}J_2^-
+\hbar\omega_2J_1^++\hbar^2\left(\frac{m^2}{2}-2r^2\right),
\label{K2sl3Add}
\end{eqnarray}
and the transformed first integral is
\begin{eqnarray}
\widetilde X=-\hbar^2(J_2^-)^2+2\hbar\omega_2J_{22}
+\hbar\omega_2-\frac{g_{2,m}^2}{\omega_2^2}.
\label{A2sl3Add}
\end{eqnarray}
The transformed second integral has the same explicit operator pattern,
\begin{align}
\widetilde{\mathscr Y}_2={}&-2\hbar^2J_{11}J_2^-
-\hbar^2J_{22}J_2^-+\frac{\hbar^2g_{2,m}}{\omega_2^2}(J_2^-)^2
+2\hbar\omega_2J_2^++\frac{\hbar\omega_2}{2}J_{12}\nonumber\\
&-\hbar^2(2r+1)J_2^--\frac{2\hbar g_{2,m}}{\omega_2}J_{22}
+\frac{\hbar g_{2,m}}{\omega_2}(2N+2r+1).
\label{B2sl3Add}
\end{align}
Hence the full second-model Sturm Hamiltonian and its two commuting integrals are likewise contained in the same $U(\mathfrak{sl}(3))$ representation.  Comparing Eqs.~(\ref{K1sl3Add})--(\ref{B1sl3Add}) with Eqs.~(\ref{K2sl3Add})--(\ref{B2sl3Add}) shows that the gauge-rotated operators have one universal projective $\mathfrak{sl}(3)$ template: the two realizations differ only through the substitutions $\omega_1\mapsto\omega_2$ and $g_1\mapsto g_{2,m}$ (together with the corresponding reduced energy shift).  This makes the common $\mathfrak{sl}(3)$ algebraization a direct consequence of the common reduced Hamiltonian (\ref{commonReducedForm}).

For fixed $E$ define
\begin{eqnarray}
\Lambda_{2,p}(E)=\frac{2}{\omega_2}[\overline E_2-\hbar\omega_2(p+1)].
\end{eqnarray}
The Sturm coupling spectrum is
\begin{eqnarray}
u_{2,p}^{(2)}(E)=\frac12[\Lambda_{2,p}(E)^2-m^2\hbar^2]
=\frac{2}{\omega_2^2}[\overline E_2-\hbar\omega_2(p+1)]^2-\frac{m^2\hbar^2}{2}.
\label{M2SturmSpecAdd}
\end{eqnarray}
The corresponding wavefunctions are
\begin{align}
\Psi_{p;n_\rho,n_z}^{(2)}={}&e^{im\phi}\rho^{\Lambda_{2,p}/\hbar}
\exp\left[-\frac{\omega_2\rho^2}{4\hbar}
-\frac{\omega_2}{2\hbar}\left(z+\frac{g_{2,m}}{\omega_2^2}\right)^2\right]\nonumber\\
&\times L_{n_\rho}^{(\Lambda_{2,p}/\hbar)}\left(\frac{\omega_2\rho^2}{2\hbar}\right)
H_{n_z}\left(\sqrt{\frac{\omega_2}{\hbar}}\left(z+\frac{g_{2,m}}{\omega_2^2}\right)\right),
\qquad n_\rho+n_z=p.
\label{M2SturmWFAdd}
\end{align}
This completes the operator-level Sturm calculation for Model II.  Combining the results of Sects.~4, 6 and 7, the algebraic organization is therefore parallel for the two systems: Models I and II each admit a two-variable $\mathfrak{sl}(3)$ algebraization of the full Sturm problem before separation and a one-variable $\mathfrak{sl}(2)$ algebraization of each parabolic separated equation.  This common Lie-algebraic pattern should not be confused with equivalence of the original magnetic Hamiltonians.  It reflects the common normal form reached after fixing the axial integral and performing the Sturm reduction.

\section{Conclusion}
We have analyzed two quantum superintegrable systems in genuine, non-vanishing magnetic fields by combining their symmetry algebras with direct solution of the Schr\"{o}dinger equation.  For both systems the two non-central conserved quantities, together with the central axial integral $Z$, generate a quadratic algebra $Q(3)$ with three generators.  Its Casimir reduces to a function of the central elements, and an explicit deformed-oscillator realization of the generators $X$, $Y$ and $W$ reduces the representation problem to a single structure function whose finite-dimensional modules give the algebraic energy branches.  The representation conditions are not used in isolation: the same energies are checked by explicit separation of variables.

In circular parabolic coordinates, both models lead to pairs of sextic singular-oscillator equations with opposite quartic coefficients and opposite separation constants.  After the behavior at the origin and infinity is factored out, these equations reduce to biconfluent-Heun form.  The first polynomial termination condition fixes the finite energy sector, while the second condition determines the separation constant.  The $\mathfrak{sl}(2)$ realization makes this second condition explicit: on $\mathcal P_N$ the Sturm operator becomes a tridiagonal $(N+1)\times(N+1)$ matrix, and its characteristic polynomial gives the allowed parabolic separation constants.  The two parabolic matrices are related by a simple spectral reflection, so the $\xi$ and $\eta$ separation constants occur with opposite signs as required by separation. 

Cylindrical coordinates provide a complementary exactly solvable description.  In each model the radial equation is of singular-oscillator/confluent-hypergeometric type and the longitudinal equation is a shifted harmonic oscillator.  The resulting Laguerre--Hermite wavefunctions reproduce the positive finite-dimensional energy branch obtained from the quadratic algebra and from the $\mathfrak{sl}(2)$ Sturm condition.  This agreement gives an internal check between the algebraic and differential-equation calculations.  The separated $\mathfrak{sl}(2)$ construction therefore provides a second algebraic route to the spectrum, whereas cylindrical separation gives a direct exactly solvable realization.  In particular, the same superintegrable system can display exact or quasi-exact solvability depending on the coordinate system and on whether the algebraization is performed before or after separation.  This interplay is especially relevant for multiseparable systems and remains comparatively little explored for Hamiltonians with magnetic fields.

We have also constructed a Sturm representation before separation.  On a fixed axial sector $Z=m\hbar$, the inverse-square coupling $u_2$ is promoted to the Sturm eigenvalue.  The integrals of motion transform together with the Hamiltonian and continue to commute with the full Sturm operator.  After introducing $x=\rho^2$, shifting the longitudinal coordinate and applying the oscillator gauge factor, the Sturm Hamiltonian and both non-central integrals belong to $U(\mathfrak{sl}(3))$ and preserve a finite triangular polynomial module.  The Frobenius exponent $r$ is directly related to the Sturm coupling $u_2$; the physical Sturm eigenspace is the level selected inside that module, and its separated Laguerre--Hermite basis yields the coupling spectrum and wavefunctions.  This provides a third algebraic structure, now associated with the symmetry algebra of the Sturm form rather than directly with the original Schr\"odinger Hamiltonian.  The fact that both models lead to this same $\mathfrak{sl}(3)$ realization is explained by their common reduced singular-oscillator normal form after fixing $Z=m\hbar$, not by an equivalence of their original magnetic fields.  More precisely, the two systems have the same reduced Hamiltonian and transformed quadratic-algebra templates after different effective parameter identifications, while the dependence of those parameters on the axial quantum number is different: Model I carries the $m$-dependence in the effective oscillator frequency and additive shift, whereas Model II carries it in the effective linear coupling.  The distinction between the classified magnetic systems is therefore retained when the fixed-$m$ sectors are reassembled into the full three-dimensional problem.

The two Lie-algebraic descriptions therefore have complementary roles.  For each system, the complete two-variable Sturm problem admits an $\mathfrak{sl}(3)$ algebraization before separation, whereas each one-variable parabolic Sturm equation admits an $\mathfrak{sl}(2)$ algebraization after separation.  The quadratic algebra $Q(3)$ remains the symmetry algebra of the original superintegrable Hamiltonian.  Taken together, these three structures connect the integrals of motion, finite-dimensional representations, separated special-function solutions and the Sturm coupling problem in a single framework.  More generally, the relation between different frameworks for solvability and superintegrability remains incompletely explored, particularly in the presence of magnetic fields.  The present examples show that distinct classified magnetic systems can nevertheless share a common reduced algebraization after symmetry reduction.

\section*{Acknowledgements}
FH was supported by PUST-RTTC-Project-2025-July-15/108. IM was supported by the Australian Research Council Future Fellowship FT180100099. 

\section*{Data availability}

\noindent The manuscript has no associated data.

\end{document}